\documentclass[aps,prb,singlecolumn]{revtex4}
\usepackage{graphicx}
\newcommand\beq{\begin{equation}}
\newcommand\eeq{\end{equation}}
\newcommand\bear{\begin{eqnarray}}
\newcommand\eear{\end{eqnarray}}

\begin{document}

\title{Lifting of Ir\{100\} reconstruction by CO adsorption: 

an \textit{ab initio} study}

\author{Prasenjit Ghosh and Shobhana Narasimhan}

\affiliation{Theoretical Sciences Unit, Jawaharlal Nehru Centre 
for Advanced Scientific Research, \\
Jakkur, Bangalore 560 064, India }

\author{Stephen J. Jenkins and David A. King}

\affiliation{Department of Chemistry, University of Cambridge,\\
Lensfield Road, Cambridge CB2 1EW, U.K.\\}

\date{\today}

\begin{abstract}

The adsorption of CO on unreconstructed and reconstructed Ir\{100\} has been 
studied, using a combination of density functional theory and thermodynamics,
to determine the relative stability of the two phases as a function of CO
coverage, temperature and pressure.  We obtain good agreement with experimental
data. At zero temperature, the (1$\times$5) reconstruction becomes less
stable than the unreconstructed (1$\times$1) surface when the CO coverage
exceeds a critical value of 0.09~ML.
The interaction between
CO molecules is found to be repulsive on the reconstructed
surface, but attractive on the unreconstructed, explaining the experimental observation of high CO coverage on growing
$(1\times 1)$ islands.  At all temperatures and pressures, we find only two possible
stable states: 0.05 ML CO c$(2 \times 2)$ overlayer on the  (1$\times$1) substrate,
and the clean (1$\times$5) reconstructed surface.
\end{abstract}

\maketitle

\newpage

\section{INTRODUCTION}

\label{intro}

Automotive exhausts contain several noxious gases, such as CO and NO, which 
need to be converted to less harmful products (e.g., CO oxidized to
CO$_2$, and NO reduced to N$_2$) before they are released into 
the atmosphere. The surfaces of metals like Pt, Ir and Rh are good 
catalysts for such reactions. A better understanding of the 
catalytic activity of these surfaces could lead to the development
OF cheaper and better catalysts, and these reactions have therefore 
been the subject of extensive experimental and theoretical studies\cite{Liu}.

In response to reduced coordination at the surface, the \{100\} surfaces of 
the face centered cubic (fcc) metals Au, Pt and 
Ir reconstruct to form a corrugated quasi-hexagonal overlayer (hex) on 
top of the square fcc substrate. While Pt\{100\}
and Au\{100\} display complex periodic patterns with large unit 
cells\cite{AuPt}, Ir\{100\} has a similar structure, but with a comparatively 
small (1$\times$5) unit cell. 
Low energy electron diffraction (LEED) experiments\cite{Hove, Lang, Bickel, Johnson} 
and density functional theory (DFT) calculations \cite{Ge} on Ir\{100\} 
have shown that the stable overlayer registry with 
respect to the substrate is in accordance with the ``two-bridge model". 
The lack of registry between overlayer and substrate atoms results 
in a significant buckling in the two topmost layers\cite{Johnson}, 
as well as a lateral shift in the second layer. Moreover, the 
reconstruction appears to go down deep, into at least the fourth 
layer\cite{Schmidt}.
Though most previous studies have focused on adsorption and reaction mechanisms on 
Pt\{100\} rather than Ir\{100\}, in this work we choose to focus on Ir\{100\},
since its smaller unit cell makes it more accessible to \textit{ab initio} 
calculations. However, the structural 
similarity of the reconstructions on Pt\{100\} and Ir\{100\} suggests that the
chemical reactions and physical processes taking 
place on the two surfaces might have similar mechanisms. 

Surface adsorbates such as CO, NO and O$_2$ are known to lift the 
reconstruction on Pt\{100\} and Ir\{100\}. For Pt\{100\}, the 
adsorbate-induced lifting of the reconstruction has been widely 
investigated\cite{Thiel, CO1, CO2, Ritter, Borg}. 
There have also been a few studies on the lifting of
reconstruction on Ir\{100\} by molecular adsorbates\cite{COIr, O2Ir, Deskins, Yeo95,Khatua}.
Measurements of heats of adsorption, on both unreconstructed and  
reconstructed Pt\{100\} and Ir\{100\}, by Hopkinson \textit{et al.} \cite{CO2},
Yeo \textit{et al.} \cite{Yeo95} and
Ali \textit{et al.} \cite{COIr}, have shown that the binding energy of CO is greater on the
unreconstructed surface than the reconstructed one; 
this provides the driving 
force for the lifting of the reconstruction upon adsorption of CO.

The catalytic oxidation of CO on metal surfaces is certainly one of the most important
catalytic processes studied in surface science. When the catalyst is Pt\{100\} or Ir\{100\},
the CO-induced lifting of the reconstruction forms a crucial component of the catalytic
cycle, since the surface alternates rapidly between being CO-rich and (after the
CO has combined with oxygen to form CO$_2$) being clean. Thus, the surface alternates
also between being unreconstructed and reconstructed, and the thermodynamics and kinetics
governing this process are of great interest if one wishes to gain a better understanding
of the catalytic cycle.

The parameters that characterize the process by which
the reconstruction is lifted are: (i) the adsorbate coverage on the metal surface,
(ii) the nucleation of unreconstructed $(1 \times 1)$ islands, and (iii) the growth rate of these
islands. In the following paragraphs, we summarize the present state of knowledge
regarding these three parameters:

(i) {\textit {Coverage}}: What is the coverage of CO on the $(1 \times 1)$ islands, and at what critical coverage
is the lifting initiated? Is the critical parameter the {\textit {global}} CO coverage,
or is it a {\textit {local}} CO coverage (that differs from the average value) that is
important? In experimental investigations of the restructuring process on Ir\{100\} 
and Pt\{100\}, it has been reported that the local CO coverage on the 
growing (1$\times$1) islands is 0.5 ML. However, the critical value of the
total CO coverage, for the onset of the transition, has been found to
be much lower than this, on both Ir\{100\} and Pt\{100\}. For example, thermal energy
atomic scattering measurements on Ir\{100\} and Pt\{100\} suggest that the
restructuring begins somewhere between 0.05 and 0.13 ML on Ir\{100\}\cite{COIr}, and 
between 0.01 and 0.03 ML on Pt\{100\}\cite{CO2}; while
electron energy loss spectroscopy measurements by 
Behm \textit{et al.}\cite{Behm} and Rutherford back scattering  
measurements by Jackman \textit{et al.}\cite{Jackman} on Pt\{100\}
have indicated that the lifting of the reconstruction 
is initiated at a critical CO 
coverage of 0.05 ML and 0.08 $\pm$ 0.05 ML 
respectively. When the total CO coverage is 0.5 ML, the entire surface 
appears to be in the (1$\times$1) phase. 

(ii) {\textit {Nucleation}}: It is not clear from the literature whether the nucleation of (1$\times$1) 
islands is homogeneous or heterogeneous. 
On the one hand, the finding of  
similar CO adsorption energies on steps and terraces reported by Hopster 
\textit{et al.}\cite{Ibach}, indicates that the nucleation may be homogeneous.
Using scanning tunneling microscopy
(STM), Ritter \textit{et al.}\cite{Ritter} have proposed that
homogeneous nucleation of the (1$\times$1)
islands takes place due to fluctuations in the density of CO molecules. 
According to them, when the islands grow beyond 
a critical size, they become stable and act as nucleation centers.
The spatial progress of the transformation occurs when the 
rate of growth of the islands becomes large compared to 
their rate of nucleation.   
On the other hand, STM studies by Borg \textit{et al.}\cite{Borg1} 
suggest that the restructuring is initiated by 
heterogeneous nucleation, the nucleation centers being step edges 
and structural irregularities disrupting the hexagonal structure 
along a direction close to the [$\overline{1}~5$]  direction of
their [N~1~;~$\overline{1}~5$] reconstruction. 
In agreement with this, molecular dynamics (MD) simulations by van Beurden 
\textit{et al.}\cite{Borg, Kramer} of the lifting of the reconstruction 
on Pt\{100\}, at CO coverages between 0.4 and 0.5 ML, 
indicate that the transformation
is heterogeneously nucleated at step edges aligned along the
[011] direction.    

(iii) {\textit {Growth rate}}: From their molecular beam experiments, Hopkinson \textit{et al.}
deduced how $r_{1\times1}$, the growth rate of the (1$\times$1) islands,
depends on the local CO coverage on the hex-surface 
($\Theta_{CO}^{\mbox{hex}}$). They obtained a non-linear variation of the
growth rate: $r_{1\times1}=k (\Theta_{CO}^{\mbox{hex}})^n\Theta_{hex}$, where $k$ is 
independent of $\Theta_{CO}^{\mbox{hex}}$, $\Theta_{hex}$ is the fraction of surface remaining in the reconstructed form, and the reaction order $n$ was found to be $4.5\pm 0.4$. 
From this, they concluded that 4 to 5 CO molecules must be involved 
{\textit {cooperatively}} in the growth
of the (1$\times$1) phase. Similar studies by Ali 
\textit{et al.}\cite{COIr} of the 
CO-induced lifting of reconstruction on Ir\{100\} also showed
a power law behavior for the growth of (1$\times$1) islands,
with $3.9\leq n\leq5.8$. Recent time-resolved reflection absorption 
infrared spectroscopy (RAIRS) results indicate a similar power law
relationship by an alternative, and very direct, technique \cite{Pratt}.

For the Pt surface, the presence of a non-linear term in the growth law appears to be
crucial to the appearance of oscillations in many catalytic 
processes, e.g., CO oxidation with NO \cite{CONO1, CONO2}, CO oxidation with 
O$_2$\cite{COO21, COO22} and NO reduction with H$_2$\cite{NOH21, NOH22}.
Under reaction conditions, the catalyst surfaces undergo transformation; these
are non-equilibrium processes. Moreover the reaction rate may not remain
constant but changes periodically or exhibits chaotic behavior. There can
also be the formation of spatial patterns on the catalyst surface\cite{Nicolis}, and these have been modelled recently within a sophisticated reaction-diffusion scheme\cite{Anghel}.
Although we are not aware of any literature concerning oscillatory reactions on 
Ir\{100\}, the experimental findings by Ali \textit{et al.} suggest that 
such oscillatory behavior may be present at temperatures above 900 K.

What is the underlying mechanism that is responsible for the non-linear growth law?
Hopkinson \textit{et al.}\cite{CO1} and Ali \textit{et al.}\cite{COIr} 
have proposed a simple mechanism for the restructuring process which involves
a cooperative phenomenon among CO molecules: due to statistical fluctuations 
of the local CO coverage on the 
hex phase, 4-5 CO molecules come together at the boundary of the growing 
(1$\times$1) domain, or at a step, and convert 8-10 Pt atoms from a hex to square 
arrangement; however it is not clear {\textit {why}} 4-5 CO molecules are needed or
precisely {\textit {how}} they cooperate.
One explanation for the `magic' value has been suggested by Passerone 
\textit {et al.}\cite{Tosatti}, who performed MD simulations on Au\{100\}, and
found that when islands/craters are formed by adsorbing/desorbing Au atoms,
they do not remain stable unless they exceed a critical size of 8-10 Au atoms.
However, it is not clear that this number will translate to Pt or Ir surfaces,
especially in the presence of CO. The restructuring of Pt\{100\} has also
been studied in more recent MD simulations by van Beurden and coworkers\cite{Borg}, 
who have found that the CO 
molecules initiate surface relaxations and lead to a shear tension between 
adjacent $\lbrack0\overline{1}1\rbrack$ rows due to the preference for a 
square rather than a hexagonal coordination sphere, and the
restructuring proceeds through the  ejection of chains 
of Pt atoms, and a rearrangement of the remaining surface atoms. Their simulations
showed no evidence of a cooperative phenomenon between 4-5 CO molecules, and the
source of the non-linear growth law remains a mystery. However, it is important to note 
that their MD simulations were done at much higher CO coverages (0.4-0.5 ML) 
than the critical coverage at which the experiments showing the fourth-order power law dependence of the rate of (1$\times$1) formation were conducted.
The mechanism for the lifting of the reconstruction is clearly not the same for low and high CO 
coverage.  At low coverage, lifting of the reconstruction occurs when random
statistical fluctuations bring about a sufficiently high local coverage
to instigate nucleation of (1$\times$1) islands; at the high total coverages
studied in the MD simulations, such high local coverage would be found
across the whole surface at all times. The experimentally-observed power law dependence stems from
the probability of bringing together 4-5 CO molecules within a sufficiently
small area at low overall coverage.

As a first step towards quantifying these arguments theoretically,
it is desirable first to understand the thermodynamics, 
i.e., at what coverage the phase transition becomes thermodynamically 
favorable. In this work, we determine this quantity through 
the study of the thermodynamics of the reconstructed and unreconstructed 
surfaces using \textit{ab initio} DFT. We then extend these results to finite
temperatures and pressures, by making use of the chemical potential. This is
useful because it brings the results into regimes where it becomes possible
to compare with experiment.

Our work is similar in spirit to a recent DFT study 
by Deskins \textit{et al.}\cite{Deskins}, who showed that the
adsorption of atomic oxygen on Pt\{100\} makes the unreconstructed 
surface more thermodynamically stable than the reconstructed one. In their work, the reconstructed
structure of Pt\{100\} was approximated by a $(1 \times 5)$ unit cell. Note however
that: (i) our focussing on Ir\{100\} enables us to use a relatively small unit cell that
corresponds to the {\textit {true}} reconstructed structure; (ii) we have
studied the adsorption of CO; (iii) details of our analysis are also 
different. 

The rest of this paper is structured as follows: Section \ref{FPdet} provides
some details about our first-principles calculations. 
Sections \ref{ucleansur} and \ref{rcleansur} contain the results of our 
calculations on the clean Ir\{100\}
surface (both unreconstructed and reconstructed), while our results
for CO adsorbed on the unreconstructed 
and reconstructed Ir\{100\} surfaces are contained in 
Sections \ref{COunrecsur} and \ref{COrecsur} respectively. We emphasize that
it is particularly important to choose k-point meshes (used for Brillouin
zone integrations) very carefully; this issue is discussed in
Section \ref{FPdet}, while some illustrative examples are presented in
\ref{COunrecsur}.
The {\it ab initio} density functional theory results feed into an analysis using the 
entropy and the chemical potential,
presented in Section \ref{thermodynamics}. Finally, we discuss the implications of our results and summarize
in Section \ref{concl}.

\section{DETAILS OF \textit{AB INITIO} CALCULATIONS} 

\label{FPdet}

It is known that the lifting of the reconstruction in the systems under
study is governed by very small
differences in the energies of the competing structures; it is therefore vital 
to do as accurate a calculation as possible. For this reason, we have chosen to perform
\textit {ab initio} calculations 
within the framework of density functional 
theory\cite{DFT} (DFT), since this is perhaps the most reliable method currently available for obtaining accurate values of ground state 
properties such as structures, surface energies and adsorption energies.

Our calculations have been performed using the CASTEP package, 
wherein the Kohn-Sham\cite{KS} equations are solved iteratively by
conjugate gradient minimization.\cite{CG}
We have used a plane wave basis set, with a cut-off of 25 Ry (340) eV, and
ultrasoft pseudopotentials. The pseudopotentials for Ir, C and O are
Ir\_OO.usp, C\_OO.usp and O\_OO.usp respectively which are provided along
with the CASTEP distribution (Version 4.2). For the exchange-correlation 
interactions, we have used the Perdew-Wang form of the generalized 
gradient approximation (GGA) \cite{Wang};
note that earlier calculations\cite{Ge} showed that gradient corrections are
essential in properly describing this surface, and that use of the
local density approximation (LDA) instead of the GGA would incorrectly predict clean Ir\{100\} to be stable against reconstruction.
Integrations over the Brillouin zone have been evaluated with a 
Monkhorst-Pack (MP) \cite{MP} mesh (further details of which are given below),
along with a Gaussian smearing function of width 0.1 eV.

In order to test the reliability of the pseudopotentials 
used in our calculations, 
we first performed calculations on bulk Ir and a CO molecule in the gas 
phase. For the former we obtain a lattice constant of 3.86 \AA, which is in 
excellent agreement with the experimental value of 
3.84 \AA, while for the latter, we obtained a C-O bond length of 
1.14 \AA  which 
also agrees very well with the experimental value of 1.13 \AA\cite{CRC}.

For surface calculations, we use a supercell consisting of a slab of six 
layers of Ir 
atoms separated by a vacuum thickness of about 10 \AA. The top four Ir layers 
are allowed to relax their positions, whereas the bottom two are fixed at 
the bulk separation (`asymmetric' slab). Further, only the top surface is allowed
to reconstruct; note that the density of atoms is different on a reconstructed
and unreconstructed surface. Thus, all such asymmetric slabs will contain one
surface that is of interest to us, while the other side consists of a bulk-truncated
(unrelaxed and unreconstructed) Ir\{100\} surface. In order to determine the surface
energy of the latter, we also perform a calulation on a `symmetric' slab, comprised
of eight layers, of which the middle two are fixed at the bulk spacing, while the top
three and bottom three layers are allowed to relax. A comparison of the symmetric
and asymmetric slabs for the unreconstructed structure enables one to determine
separately the surface energies of a relaxed and bulk-truncated $(1\times1)$ surface.

\begin{figure}[p]

\centering

\includegraphics[scale=0.40]{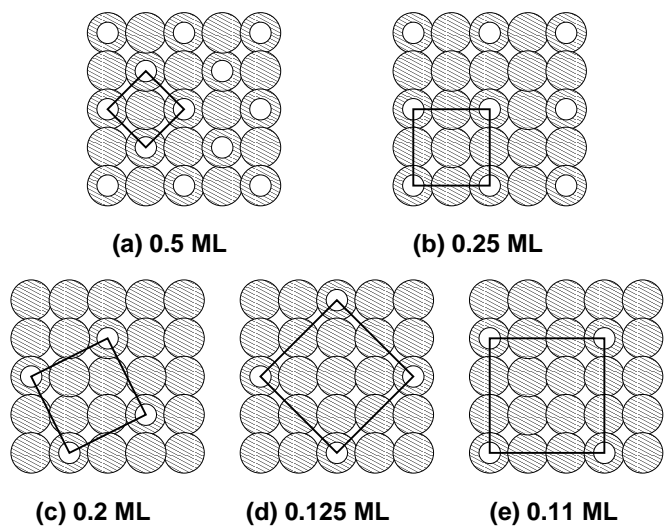}

\caption{Schematic top views of the surface unit cells for different 
CO coverages. The large shaded circles 
and the small white ones denote Ir atoms on the topmost layer and
CO molecules respectively. The unit cells
are indicated by solid black lines: (a)$\sqrt{2}\times\sqrt{2}$, (b) 2$\times$2,
(c)$\sqrt{5}\times\sqrt{5}$, (d) 2$\sqrt{2}\times$2$\sqrt{2}$ and 
(e) 3$\times$3. }

\label{unrecCO}

\end{figure}

The size of the surface unit cell used 
depends upon whether whether we are looking at an unreconstructed surface or
a reconstructed one, and what CO coverage we are considering. Calculations for the
clean unreconstructed surface were carried out using both $(1 \times 1)$ and
$(1 \times 5)$ cells; the results obtained with the two were almost identical.
For the clean reconstructed surface, we use a $(1 \times 5)$ cell. 
To study CO adsorption on the unreconstructed and reconstructed surface we
adsorb CO on only one side (the side which we are allowing to relax) of the 
'asymmetric slab'. For adsorption of CO on the unreconstructed surface,
we consider coverages of 0.11, 0.125, 0.2, 0.25
and 0.5 ML. The unit cells used for these calculations are shown in 
Fig. \ref{unrecCO}; note that (i) all the unit cells are square, i.e., 
the distance between adjacent CO molecules is the same in both directions, 
and (ii) the cells for different coverages are not necessarily commensurate with one another.
Finally, for studying adsorption of CO on the reconstructed surface,
we use a $(2 \times 5)$ cell, within which we consider CO coverages of
0.1, 0.2 and 0.6 ML. (Note that all CO coverages in this paper are given with 
respect to the density of atoms in the topmost layer of the unreconstructed 
surface.)

An important consideration is the choice of k-point meshes for Brillouin zone (BZ) sampling.
For the surface cells, we have used Monkhorst-Pack meshes, of the form $(n_1 \times n_2 \times 1)$,
where $n_1$ and $n_2$ determine the fineness of the mesh.
This issue becomes particularly crucial in the present problem, since we are interested in
computing energy differences that are comparable to the errors introduced by incorrect (unconverged)
BZ sampling. Such errors can be reduced by using a k-point mesh that is as fine as possible,
while keeping computational feasibility in mind. We have found that, as expected,
the convergence is faster when one uses a mesh that does not include high symmetry
points [the Brillouin zone center (BZC) and the k-points on the edges of the BZ];
this corresponds to choosing $n_1$ and $n_2$ to be even numbers. 
Examples of this are presented below in Section \ref{COrecsur}.

The specific choices made for
$n_1$ and $n_2$ for the various surface cells used in the present paper are given in Table \ref{kpts};
we emphasize that convergence with respect to k-point sampling has been carefully established for
all the cases. Note also that computing surface energies and adsorption energies requires comparing
the total energies of two systems (bulk and clean surface, or clean and covered surface); in such
cases, we are careful to make sure that the unit cells and k-point meshes used for the two systems
are either identical or related by folding, thereby further reducing the errors introduced by
incomplete k-point sampling.  

\section{RESULTS}

\label{results} 

\subsection{Unreconstructed Clean Surface}

\label{ucleansur}

Using the asymmetric six-layer slab described above, we find that the first 
interlayer distance $d_{12}$ is contracted (with respect to the bulk 
interlayer separation) by 
6.14\%. This result compares well with that obtained 
from 
\textit{ab initio} calculations by Ge \textit{et al.}\cite{Ge}(6.5\%), 
but is larger than the reported experimental value ($\sim$ 3.6\%)\cite{Heinz}.
Upon comparing the total energy of this asymmetric slab with that of the bulk structure, 
we find that the \textit {sum} of the surface energies
of the two surfaces (one relaxed and the other unrelaxed) is 3.01 eV per $(1\times1)$ 
area. Next, using
the symmetric eight-layer slab, which possesses two relaxed surfaces, we determine
the energy of the relaxed unreconstructed surface, $\Gamma^{rel}_{1 \times 1}$, to be 1.50
eV per $(1 \times 1)$ area, while the energy of the unrelaxed (bulk-truncated) unreconstructed
surface, $\Gamma^{unrel}_{1 \times 1}$, is 1.51 eV per  $(1 \times 1)$ area. This latter
quantity is subtracted out when determining surface energies for slabs that are reconstructed
and/or have CO adsorbed on only one side.

\subsection{Reconstructed Clean Surface}

\label{rcleansur}

\begin{figure}[p]

\centering

\includegraphics[scale=0.40]{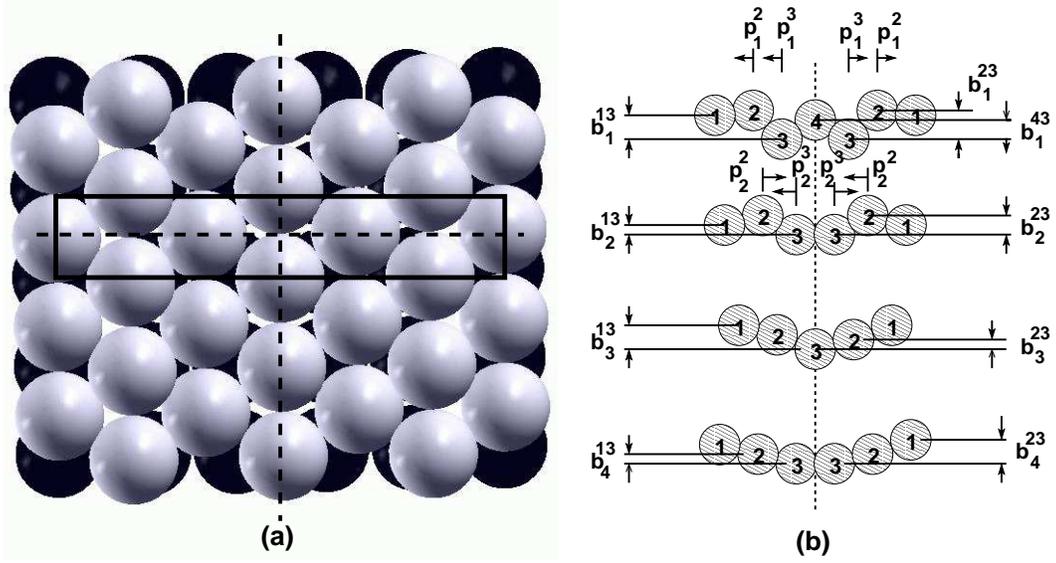}

\caption{(a) Exact top view and (b) schematic side view of the reconstructed 
Ir\{100\} surface. In (a) the grey spheres are atoms in the topmost layer, 
whereas the black ones are atoms in the second layer. The surface
unit cell is demarcated by the solid black rectangle. The horizontal and
dashed lines denote the planes of reflection  of the unit cell. In (b),
$b^{i3}_n$ is the distance along the z-direction between atoms $i$ and $3$
in the $n^{th}$ layer and $p$ denotes the lateral shift of the atoms from
their bulk truncated position. The arrows denote the direction 
of the shift. This is a schematic diagram. The actual values of different
parameters are listed in Table \ref{clrecp}.}


\label{clrec}

\end{figure}

In the reconstructed surface, the topmost layer forms a quasi-hexagonal layer on top of the square
substrate, as can be seen in Fig. \ref{clrec}(a). The surface unit cell is $(1 \times 5)$. 
The reconstruction results in  buckling within the layers, as well as lateral
shifts of the atoms with respect to their bulk-truncated positions. The parameters used to specify
these structural rearrangements are indicated in Fig. \ref{clrec}(b), and the values we obtain for
them are presented in Table \ref{clrecp}. It can be seen that
our results are in excellent agreement with
those obtained from LEED and STM
by Schmidt \textit{et al.}\cite{Schmidt}. In accordance with their 
observations, we find that in addition to the reconstruction of the topmost
layer, there are significant lateral shifts and buckling in the three layers below.
The only (minor) difference between our results and theirs is that
our calculations yield a very small
lateral shift of the third atom in the second layer ($p_2^3$) in a direction opposite
to that determined by them. 

The value we obtain for the surface energy of the reconstructed surface is
$\Gamma^{rel}_{1 \times 5}$ = 1.45 eV per $(1 \times 1)$ area, and we thus correctly 
obtain the result that the clean surface would prefer to reconstruct.  Note also that the energy of 
reconstruction for Ir\{100\} is very small, viz. 0.05 eV/(1$\times$1) area. This is
in reasonably good agreement with the value of 0.07 eV/(1$\times$1) area
obtained by Ge \textit{et al.}. The slight 
difference between our results and theirs presumably arise from the 
use of slabs of different thicknesses and different k-point meshes.  
(Note that the energy difference here is somewhat smaller than the 
value of 0.21~eV/(1$\times$1) area found experimentally for the Pt\{100\}
surface \cite{Yeo95}).

\subsection{CO on unreconstructed Ir\{100\}}

\label{COunrecsur}

\begin{figure}[p]

\centering

\includegraphics[scale=0.40]{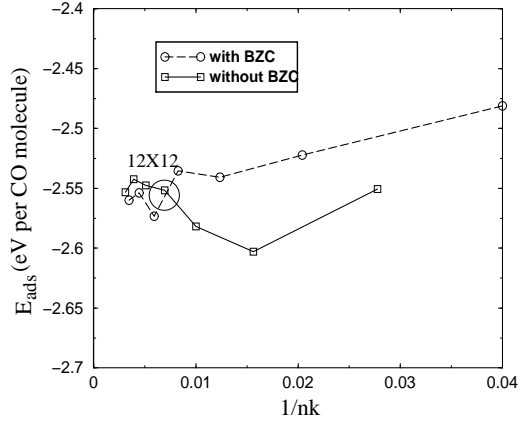}

\caption{Convergence of adsorption energy ($E_{ads}$) of CO at 0.5 ML on the
atop site of the unreconstructed surface, with respect to the k-points used
for $\sqrt{2}\times\sqrt{2}$ unit cell. Monkhorst-Pack meshes of the form 
($n_1\times n_2\times 1$) were used, meshes that include/donot include
the BZC have odd/even values of $n_1=n_2$. For $n_1=n_2=$ 5, 6, 7, 8, 9, 10,
11, 12, 13, 14, 15, 16, 17, and 18 the number of points ``nk" in the
whole BZ are 25, 36, 49, 64, 81, 100, 121, 144, 169, 196, 225, 256, 289 and
324 respectively. Note that meshes that do not include the BZC (solid line) 
converge faster than those that include it (dashed line). }  

\label{kptconv}

\end{figure}

As discussed in Sections \ref{intro} and \ref{FPdet},
the choice of a proper k-point
mesh is very crucial for our calculations. Even values of $n_1$ and $n_2$
lead to better sampling of the BZ and quicker convergence of the adsorption
energy. This fact becomes evident on inspecting Fig. \ref{kptconv},
where we have shown how the adsorption energy, for CO at 0.5 ML and using
a  $(\sqrt{2}\times\sqrt{2})$ unit cell, converges as a function of k-point
sampling. Accordingly, we use a $(12 \times 12 \times 1)$ mesh for this
particular unit cell; similar checks were performed for other coverages and
cells.

Experiments and previous theoretical calculations\cite{Titmuss} indicate that
the atop site is the most probable site for CO adsorption. To verify this, and
compare our results with previous DFT calculations and LEED measurements,\cite{Titmuss}, 
we calculate the geometry and 
adsorption energies of CO molecules occupying hollow, bridge and atop sites 
within a (2$\times$2) surface unit cell, and at 0.5 and 0.25 ML CO coverages.
The adsorption energy per CO molecule, $E_{ads}$, is given by:
\beq
E_{ads}=\frac{E_{slab+CO}-E_{slab}-n_{CO}E_{CO}}{n_{CO}},
\eeq 
\noindent where $E_{slab+CO}$, $E_{slab}$ and $E_{CO}$ are the the total energies
of the slab with CO adsorbed on it, the clean slab, and a CO molecule in the
gas phase, respectively, while $n_{CO}$
is the number of CO molecules adsorbed per surface unit cell. 
The results obtained by us for $E_{ads}$, for different sites at the two coverages
considered, are summarized in Table \ref{compCOse}. Although there 
are slight numerical differences between the adsorption energies obtained 
from our calculations and those of Titmuss \textit{et al.}\cite{Titmuss}, 
both sets of calculations predict that CO molecules adsorb on the atop site. 
The differences in the numerical values of the adsorption energy obtained 
from the two calculations presumably arise from the use of slabs of 
different sizes and different k-point mesh. The 
structural parameters are also in good agreement with both the LEED
measurements and theoretical calculations. 

We go on to study the variation of the adsorption energy of CO on the atop site
as a function of CO coverage. The results are summarized in the second column
of Table \ref{adssur}. 
The difference in adsorption energy between 0.5 ML and 0.25 ML CO coverage 
is 0.04 eV. For lower coverage, ($\Theta \leq 0.25$ ML) the adsorption energy
 is more or less constant ($\sim$ 2.51 eV). The slight variation in the 
numerical values of E$_{ads}$ is most likely due to numerical errors
that arise because the k-point 
meshes used for different surface unit cells are not exactly commensurate. 
At lower coverage, the distance between CO molecules increases, 
decreasing the interaction between them. At sufficiently low coverage, 
the CO molecules are so far apart that they do not interact with one 
another, resulting in a constant value of the adsorption energy.
The increase in E$_{ads}$ with increase in CO coverage indicates 
the presence of \textit{very weak} attractive interactions between 
nearest-neighbor CO molecules on the unreconstructed surface. 

In addition to the CO adsorption energies, 
we also calculate the ``surface energies" ($\Gamma_s$) at different CO 
coverages. These ``surface energies" will be used below as a measure of 
the stability of the reconstructed and unreconstructed surfaces. We define the 
``surface energy" of the CO covered surface ($\Gamma_s$) as:

\beq
\label{surfen_CO}
\Gamma_s=\frac{E_{slab+CO}-n_{Ir}E_{bulk}-n_{CO}E_{CO}}{N_s}-\Gamma_s^{unrel},
\eeq

\noindent where $N_s$ is the ratio of the area of the surface
unit cell to that of the (1$\times$1) cell, and the
superscript $unrel$ represents the unrelaxed and unreconstructed lower surface of the slab.

\subsection{CO on reconstructed Ir\{100\}}

\label{COrecsur}

\begin{figure}[p]

\centering

\includegraphics[scale=0.65]{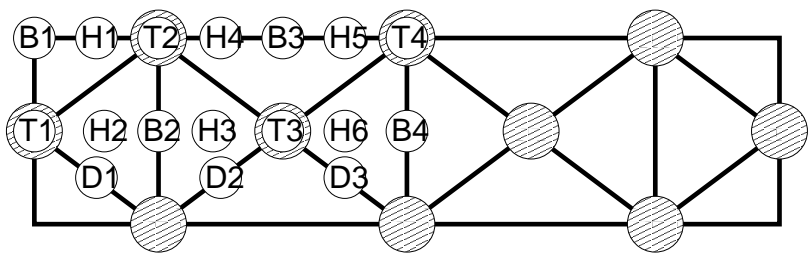}

\caption{Schematic diagram showing different possible CO adsorption sites on 
the reconstructed 
Ir \{100\} surface (top view). Shaded circles are atoms in the topmost Ir layer.
B, D, H and T denote cross bridge, diagonal 
bridge, hex and atop sites respectively. The sites are labelled according to
the nomenclature used by Deskins \textit{et al.}\cite{Deskins}. Note that
T1, T2, T3 and T4 correspond to sites atop the atoms labelled 1, 2, 3 and
4 in Fig. \ref{clrec}.}

\label{site0.1}

\end{figure}

CO adsorption on the reconstructed surface has been studied for coverages of 0.1, 0.2 
and 0.6 ML. In all the cases, we use a (2$\times$5) supercell,
so as to minimize the interaction between periodic images of CO molecules in 
adjacent supercells. Since the 
reconstructed surface is quasi-hexagonal in nature, there are many
possible adsorption sites. 
The different adsorption sites which we consider for CO adsorption at 0.1 ML 
coverage are shown in Fig. \ref{site0.1}. Our result for the 
adsorption energies for these different sites are given in
table \ref{CO_rec_ads}.
Of all the different possibilities, we find that the ``atop3" site (T3)
is the most probable one. While it is not surprising that CO prefers an
atop site, it is somewhat unexpected that the most favoured site is
atop the Ir atom that is lies lowest within the buckled surface
layer, rather than the Ir atom that protrudes. We intend to address the
origin of this very surprising result in a future publication.

\begin{figure}[p]

\centering

\includegraphics[scale=0.65]{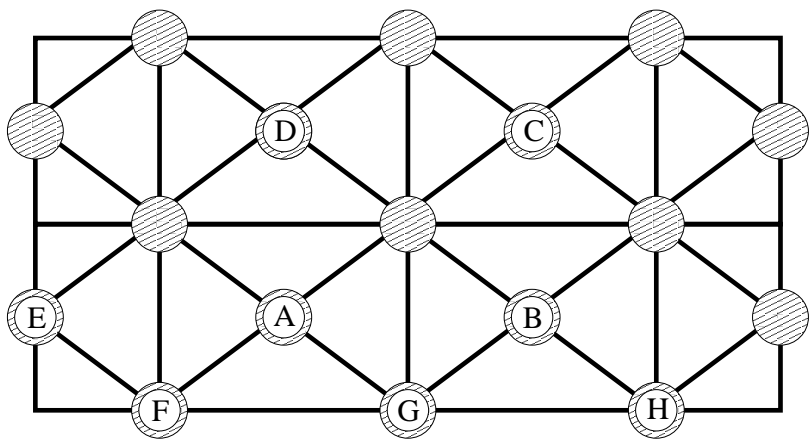}   

\caption{Schematic diagram showing different possible CO adsorption sites, within a $(2 \times 5)$ supercell, on 
the reconstructed Ir \{100\} surface (top view) at 0.2 ML and 0.6 ML CO
coverage. Shaded circles are atoms in the topmost Ir layer.
For 0.2 ML coverage the adsorption sites are A and B, A and C,
and A and D. Sites E, F, A, G, B and H
are the sites for CO adsorption at 0.6 ML CO coverage. Note that the positioning
of the lateral boundaries for the supercell follows the same convention as used
in Fig. 6; i.e., E, F and A correspond to T1, T2 and T3 sites, etc.}

\label{0.2rec-ads-site}

\end{figure}

For a coverage of 0.2 ML, we have to adsorb two CO molecules in the
(2$\times$5) cell. Since CO clearly prefers to adsorb on the ``atop3" (T3) sites,
we choose a combination of two T3 sites. There are three inequivalent
combinations of two T3 sites in the (2$\times$5) cell,
namely, A and B, A and C, and A and D (see Fig. \ref{0.2rec-ads-site}).
Our results for the CO adsorption energies at these different sites are listed in 
Table \ref{CO_rec_ads}.
Out of these three possible combination of sites, we find similar adsorption
energies for the AB and AC combinations. The fact that the AD combination
is disfavored suggests that CO molecules on this surface would prefer not to
sit too close to each other.
Comparing the adsorption energies for coverages of 0.1 and 0.2 ML,
we find that the interaction between CO molecules on the
reconstructed surface is repulsive in nature, in contrast to the 
\textit{very weak} attractive interaction on the unreconstructed surface.
For a coverage of 0.6 ML,
we need to adsorb six CO molecules per (2$\times$5) cell.
Since the CO molecules interact repulsively on the
unreconstructed surface, they will tend to spread out uniformly at
low temperatures. 
Hence we assume that at a coverage of 0.6 ML, CO will adsorb at every atop site, as shown
in Fig. \ref{0.2rec-ads-site}.

\begin{figure}[p]

\centering

\includegraphics[scale=0.65]{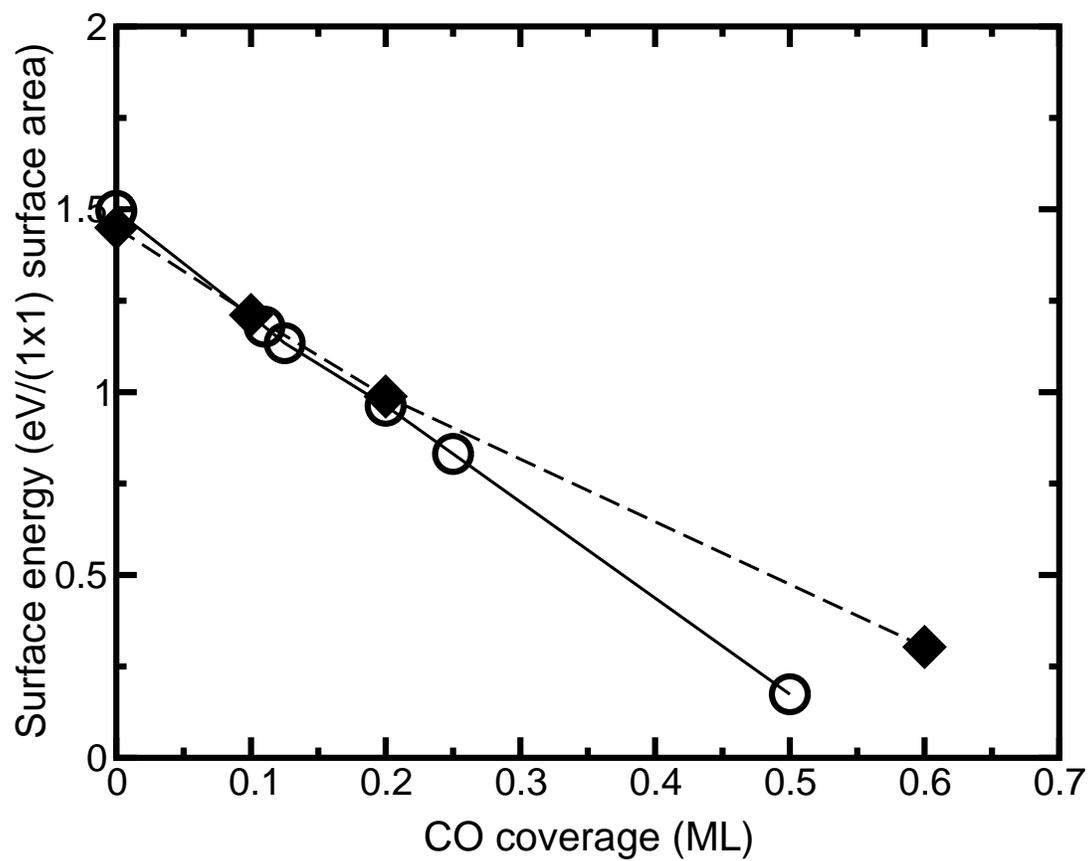}

\caption{Relative stability, at $T$ = 0 K,  of the reconstructed and the unreconstructed 
surfaces as a function of CO coverage. The dashed line and the solid line denote the 
surface energy for the reconstructed and unreconstructed surface respectively. The two lines cross in the neighborhood of 0.09 ML.}

\label{compa}

\end{figure}

The variation, with coverage,  of the ``surface energies" of the reconstructed and 
unreconstructed surfaces (computed using
Equation. (\ref{surfen_CO})), is plotted in Fig. \ref{compa}. At zero CO
coverage, the reconstructed (1$\times$5) surface is energetically more
stable than the unreconstructed (1$\times$1). 
However, at 0.1 ML coverage the unreconstructed surface is more stable
than the 
reconstructed one by about 0.03 eV per (1$\times$1) area. The crossover between
the stability of the two surfaces takes place at around 0.09 ML CO coverage. 
Thus adsorption of CO  switches the stability of the
Ir\{100\} surfaces at a very low total coverage, 
that appears to be in very good agreement with the experimentally 
reported values of Hopkinson \textit{et al.}. However, we note that
this analysis has been done at conditions corresponding to zero
temperature and pressure. In the next section, we extend this results to
finite temperatures and pressures, by performing a thermodynamic
analysis.

\subsection{Thermodynamic Analysis}

\label{thermodynamics}

In order to account for the effect of varying gas-phase temperature
and pressure upon the surface configuration, we apply concepts from
classical thermodynamics. The free energy ($F$) of a system gives a
measure of the stability of the system. For a multi-species system $F$ is given by:

\beq
\label{freeenergy}
F=E+PV-TS-\sum_{i}n_i\mu_i,
\eeq

\noindent where $P$, $V$, $T$ and $S$ are the pressure, volume,
temperature and entropy, respectively. $n_i$
and $\mu_i$ are, respectively, the number of molecules and the
chemical potential
of species $i$ at the surface. $E$ is the internal energy of the system 
and is obtained from the DFT calculations.

In order to apply these concepts we need to have a clear demarcation 
between that which we consider to be a part of the ``system" and that 
which we take to be its ``surroundings". For our case this can be achieved 
in two ways:

(i) \textit{entropy viewpoint}, in which the ``system" is considered to 
include the surface,
the adsorbed molecules on the surface and the gas above the surface; the 
``surroundings" include only the world beyond some external gas container.
Taking this view, the number of particles in the system is now fixed but
the entropy is
considerably large, since it includes large translational and rotational
contributions from the gas molecules which are not adsorbed on the surface. 
Hence Equation (\ref{freeenergy}) is modified as:

\beq
\label{f_entropy}
F^S=E+PV-TS.
\eeq 

\noindent From this viewpoint, the changes in the configuration of the surface
are driven by massive changes in entropy between molecules on the surface and
in the gas phase; net adsorption (desorption) occurs when the enthalpy of
adsorption is greater (less) than this entropy difference. 

(ii) \textit{chemical potential viewpoint}, in which the ``system" consists 
of 
the surface and any molecules adsorbed on the surface. The rest of the world
beyond then constitute the ``surroundings". Now to a fair approximation the 
entropy of the system is more or less negligible, since it consists only of
vibrational contributions. On the other hand, the number of particles in the 
system is variable, so the chemical potential of the gas molecules must be
included. The thermodynamic potential $F$ now becomes:

\beq
\label{f_mu}
F^{\mu}=E+PV-\sum_in_i\mu_i.
\eeq 

\noindent The changes in surface configuration are driven by the need to 
maintain
the equality of surface and gas-phase chemical potentials; net adsorption
(desorption) occurs when the gas-phase chemical potential is 
instantaneously greater (less) than the surface chemical potential.
Note that from both the approaches one should get the same results.

\subsubsection{Entropy Viewpoint}

\label{entropy}

Taking the viewpoint that our thermodynamic system includes the surface, adsorbed
molecules and the gas above it, we calculate the necessary free energy that must be minimized. This is achieved by defining free energies relative to that
of the clean unreconstructed surface:

\beq
\label{deltafs}
\delta F^S=\frac
{F^S_{slab+CO}-F^S_{clean(1\times1)}-\Delta n_{Ir}F^S_{bulk}-n_{CO}F^S_{CO}}
{N_s},
\eeq

\noindent where $F^S_{clean(1\times1)}$ and $F^S_{slab+CO}$ have been 
evaluated 
for the same unit cell, thus mitigating any systematic errors that may 
arise due to different k-point sampling. To evaluate the free energy of the
solid phases, slab with CO molecules adsorbed on it ($F^S_{slab+CO}$) and 
for the clean one ($F^S_{clean(1\times1)}$) we ignore the effects of $TS$
because typically the entropy of the solid phase is negligible compared to 
that of the gas phase. For the solid phases we ignore the effects of $PV$
and approximate it for CO in the gas phase, by assuming that the gas phase 
obeys the ideal gas equations. With these approximations the free energy of 
the solid
phases equals the internal energy obtained from DFT calculations.
The third term in Equation (\ref{deltafs}) contains
the free energy per atom in the Ir bulk ($F^S_{bulk}$) and the difference
in the number of Ir atoms ($\Delta n_{Ir}$) between the DFT calculations 
used to obtain the first two terms. In the last term, $F^S_{CO}$ is the free 
energy of the CO molecules in the gas phase. The last term accounts for the 
decrease in entropy of the system due to the adsorption of CO molecules
from the gas phase.
$N_s$ represents the number of
(1$\times$1) cells over which the value of $F^S$ has been calculated. 
By construction, therefore, $\delta F^S$
for the clean reconstructed surface is the negative of the reconstruction
energy, while for the clean unreconstructed surface $\delta F^S=0$.
The lowest value of $\delta F^S$ under any temperature and pressure conditions
indicates the most thermodynamically stable surface configuration,
$\delta F^S_{min}$.

In order to evaluate $F^S_{CO}$ we need to know the value of $S$ for the 
gas phase at different temperatures and pressures. The values of $S$ at a 
pressure of 1 bar and for a temperature range of 298.15 to 1500 K can be 
obtained from the CRC handbook\cite{CRC}. We then derive the variation of $S$
with pressure and temperature from the following formula:

\beq
S(T,P)=S(T,P^0)-R  \rm{ln}(P/P^0).
\eeq

\begin{figure}[p]

\centering

\includegraphics[scale=0.65]{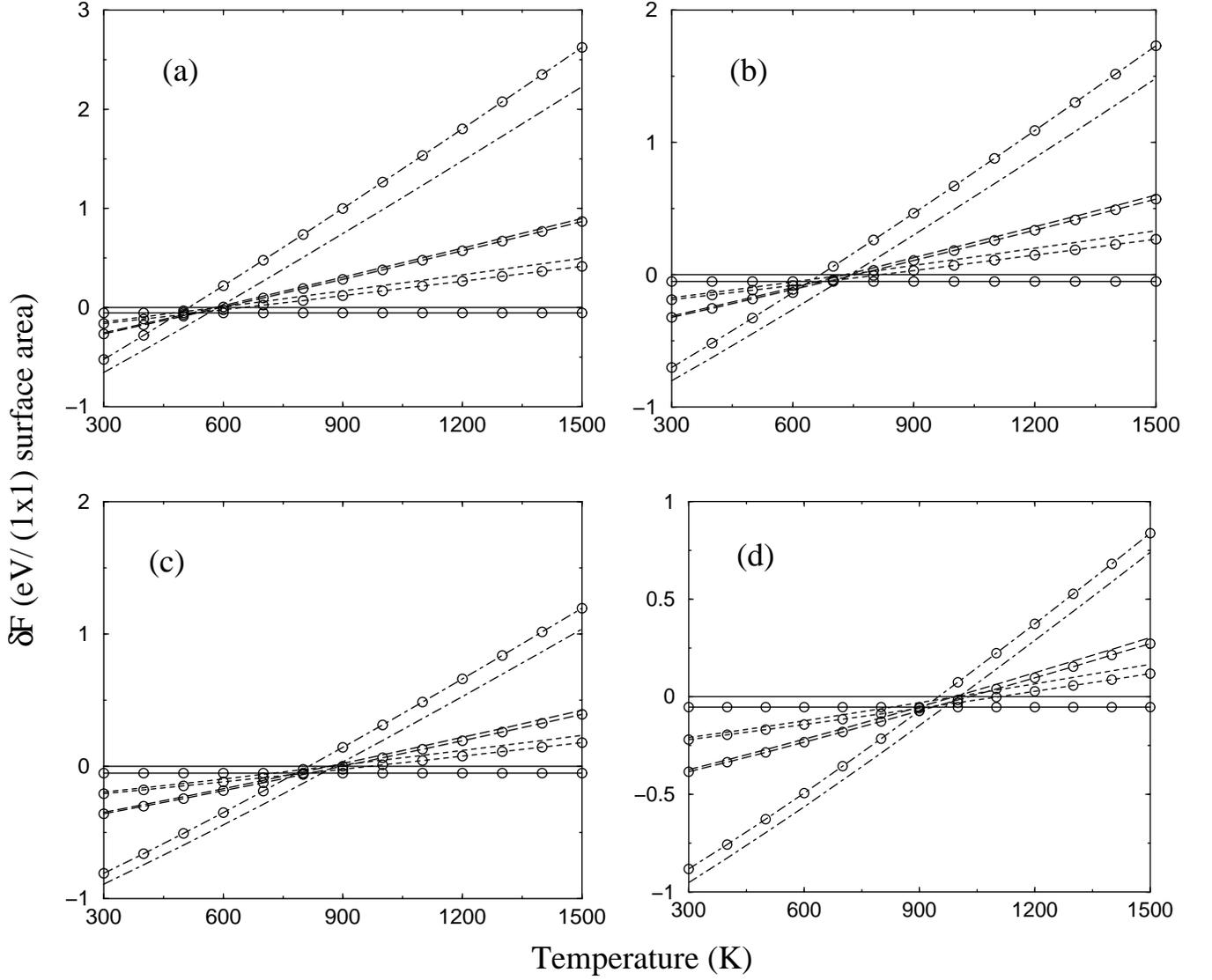}

\caption{Free energies relative to that of the clean unreconstructed surface ($\delta F$)
as a function of temperature at different pressures:
(a) $10^{-8}$ mbar, (b) $10^{-3}$ mbar, (c) $1$ mbar and (d) $10^{2}$ mbar.
The lines without any symbols on them denote the unreconstructed surface
and those with open circles denote the reconstructed one. The solid line,
the short-dashed line, and the long-dashed line 
represent  surfaces at a coverage of 0, 0.1 and 0.2 ML respectively. The dot-dashed 
line and dot-dashed line with open circles represent the 
unreconstructed surface at 0.5 ML CO coverage and the reconstructed surface
at 0.6 ML CO coverage respectively.
Note that we have dropped the superscripts $S$ and $\mu$ from $F$ since both the
approaches (as described in the text) produce the same results.
It can be seen that in all the cases, there are only two stable phases, viz., the dot-dashed
line (0.5 ML CO on unreconstructed surface) and the solid black line with circles
(clean reconstructed surface).} 

\label{deltaf_min}

\end{figure}

\begin{figure}[p]

\centering

\includegraphics[scale=0.65]{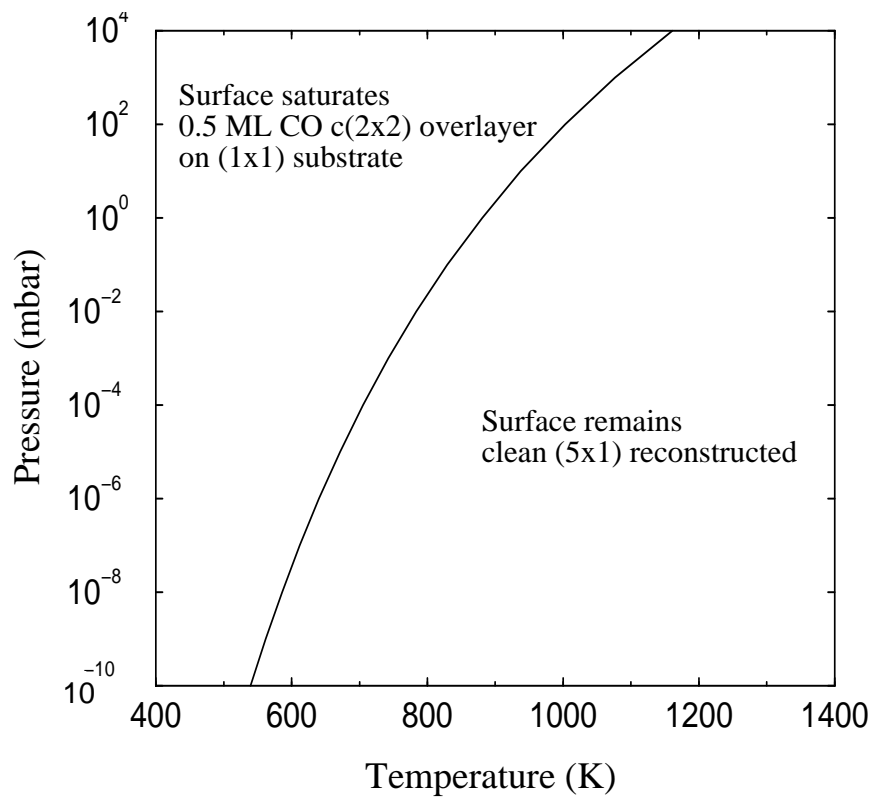}

\caption{Temperature-pressure phase diagram for CO on Ir\{100\}. The black curve dividing
the two phases gives the critical temperature at which the crossover in stability of the
two phases occurs, as a function of pressure.} 

\label{isostere}

\end{figure}

\noindent$P^0$(=1 bar) is the reference pressure and $R$ is 
the universal gas constant.
In Fig. \ref{deltaf_min}, we plot a series of lines calculated at a variety
of pressures, each plotting the value of $\delta F^S$ as a function of
temperature.
It is apparent that for all temperatures and pressures considered,
there are only two possible configurations of the surface that are 
thermodynamically stable, the clean unreconstructed (1$\times$5) surface
and the 0.5 ML c(2$\times$2) CO overlayer on the unreconstructed substrate.
The critical temperature at which the crossover in the stability occurs 
varies as a function of pressure and is depicted in Fig. \ref{isostere}.

\subsubsection{Chemical Potential Viewpoint}

In order to evaluate the relative stability of different surface configurations
from the chemical potential viewpoint, we evaluate the free energy as given 
in by Equation (\ref{f_mu}).
The chemical potential can be obtained by making use of its relationship
to the enthalpy ($H$) and entropy:

\beq
\mu_i=\frac {1}{n_i}(H-TS).
\eeq

\noindent The values of entropy at standard pressure can be extracted from
the CRC Handbook\cite{CRC}.
The CRC Handbook lists only the change in $H$ at $P^0$ in going from 
absolute zero
temperature to a variety of finite temperatures
$(\Delta H(0\rightarrow T, P^0)$. Therefore $H(T,P^0)$
can be extracted using the following relation:

\beq
H(T,P^0)=H(0,P^0)+\Delta H(0\rightarrow T, P^0)
\eeq

\noindent where $H(0,P^0)$ can be obtained from DFT calculations. The variation 
of $\mu_i$ with pressure is computed as:

\beq
\mu_i(T,P)=\mu_i(T,P^0)+k_BTln(P/P^0),
\eeq

\noindent where $k_B$ is the Boltzmann constant. Again we evaluate 
$\delta F^{\mu}$ the same way as in Equation (\ref{deltafs}) and plot the same
quantities
as evaluated in the Section \ref{entropy}. The results obtained via the 
chemical potential approach are exactly the same as those obtained via
the entropy 
approach, plotted in Fig. \ref{deltaf_min} and \ref{isostere}.

\section{Discussion and Summary}

\label{concl}

As in previous calculations, we find that Ir\{100\}
is reconstructed; the stabilization energy being very small, viz.
0.05 eV per (1$\times$1) area. Note that this is comparable to the errors
typically introduced by incorrect k-point sampling, 
emphasizing the need for careful calculations. The structure
of the reconstructed surface obtained by us is in excellent agreement
with experiment. While the increase in density is confined to the
topmost layer, the three layers below also display marked buckling
and lateral shifts.

However, as more and more CO is adsorbed on the surface, there is a reversal
in phase stability.
Our calculations show that at zero temperature and pressure, when CO is
adsorbed on Ir\{100\}, the relative stability of the reconstructed and unreconstructed phases is reversed
at a very low total CO coverage of 0.09 ML. This shows that the
lifting of the reconstruction is thermodynamically favoured at a low {\it local} coverage.
The need for statistical fluctuations to bring 4-5 molecules together
to trigger the lifting of the reconstruction is kinetic in origin,
rather than thermodynamic.

On the unreconstructed Ir\{100\} surface, CO adsorbs at atop sites, and
there is a \textit{very weak} attractive interaction between CO molecules.
Note that in this respect Ir\{100\} appears to behave differently from 
Pt\{100\},
where there is a repulsive interaction between CO molecules\cite{Behm, Yeo}.
The attractive nature of the CO-CO interactions on
the unreconstructed (1$\times$1) surface suggests that at sufficiently low temperature, CO molecules 
would tend to cluster together on being adsorbed on the fully unreconstructed surface in
order to maximize their coverage, until a saturation coverage of 0.5 ML is 
achieved. The weakness of the interaction, however, means that such clustering
may not easily be observed in experiments performed at moderate temperature.
In contrast, the interaction between CO molecules
adsorbed on the reconstructed (1$\times$5) surface is repulsive, and they tend to spread out
uniformly over the surface. Including the cost
of lifting the reconstruction, the adsorption energy of CO molecules at
any total coverage to form an island of 0.5 ML local coverage on a corresponding patch of (1$\times$1) substrate (2.45 eV per CO molecule)
is still higher than the adsorption energy of CO molecules at any coverage on 
the unreconstructed surface. This suggests that at \textit{any} total coverage,
the molecules will prefer to cluster into islands of local 0.5 ML coverage
on patches of the unreconstructed (1$\times$1) substrate, rather than
remain spread out on the reconstructed (1$\times$5) substrate. 
This bolsters 
the interpretation of 0.5 ML \textit{local} CO coverage on the 
unreconstructed (1$\times$1) islands by Ali \textit{et al.} even at low 
\textit{total} CO coverage. The higher heat of adsorption and the attractive
nature of the interaction between CO molecules on the (1$\times$1)
unreconstructed surface provide a driving force to lift the reconstruction.
For both Ir\{100\} (where the interaction between CO molecules on the unreconstructed 
surface is weakly attractive) and Pt\{100\} (where this interaction is weakly repulsive),
the main effect is that the heat of adsorption on the unreconstructed surface
far exceeds that on the reconstructed surface. As a result, the reconstruction lifting 
phenomenon remains similar on both Ir\{100\} and Pt\{100\}.
However we donot understand the driving force behind the shift of the
weakly attractive interaction between CO molecules on the unreconstructed
surface to a repulsive nature on the reconstructed one.

Our thermodynamic analysis extends the results obtained 
from DFT calculations to a range of pressures and temperatures. For the
entire temperature and pressure range considered by us, there are only
two thermodynamically stable
configurations, namely 0.5 ML of CO on the unreconstructed substrate
in a c(2$\times$2) structure, and the clean 
reconstructed substrate. The phase diagram shows that a small change in 
temperature and/or pressure can result in a very large change in coverage.
From experiments, we know that the lifting of the reconstruction sets
in at 490 K at very low pressures ($10^{-10}$-$10^{-07}$ mbar). From
our phase diagram we find that at similar pressure ranges,
the transition temperature lies between 550 and 600 K. Thus our
results from thermodynamic analysis are reasonably consistent with experimental
observations. The slight discrepancies may have arisen from our use of the ideal
gas equation of state to describe the gas phase.
In practical applications such as in catalytic converters in vehicles,
the reactions take place at temperatures above about 750 K and at $\sim$ 10 mbar
pressure. According to our phase diagram, such temperatures and pressures
lie very close to the transition line. Hence slight changes in
the reaction conditions may lead to drastic changes in the surface,
which in turn will affect the rate of chemical reactions.

To conclude, we have investigated the thermodynamic stability of 
unreconstructed and reconstructed phases of the Ir\{100\} surface,
in the presence of CO. Though our results for the critical coverage, etc.,
are in good agreement with experiment, it still remains intriguing to
speculate about the role played by kinetic factors
in the restructuring process. In particular, the origin of the nonlinear growth
law observed in experiments is still a puzzle; we hope that future work on the
kinetics of the restructuring mechanism will shed light on this.

One of us (P.G.) acknowledges CSIR, India for a research scholarship. The
Royal Society is also thanked for a University Research Fellowship (S.J.J.).
We are grateful to Dr. Stephanie Pratt for helpful discussions.

\newpage

\begin{table}[h]
\baselineskip 24pt
\caption{ Parameters for the Monkhorst-Pack k-point meshes used for different surface cells.}

\begin{center}

\begin{tabular}{c|c}

\hline

Surface unit cell& k-point mesh \\

\hline

1$\times$1& 10$\times$10\\

$\sqrt{2}\times\sqrt{2}$ & 12$\times$12\\

2$\times$2& 8$\times$8\\

$\sqrt{5}\times\sqrt{5}$&8$\times$8\\

2$\sqrt{2}\times$2$\sqrt{2}$ & 6$\times$6\\

3$\times$3 & 6$\times$6\\

1$\times$5 & 10$\times$2\\

2$\times$5 & 5$\times$2\\

\hline

\end{tabular}

\end{center}

\label{kpts}

\end{table}

\begin{table}

\caption{Structure of Ir\{100\}-{1$\times$5} described by the parameters defined
in Fig. \ref{clrec}. $\overline{d}_{ik}$'s denote average interlayer spacings 
between the centre of mass planes of the layers, while $d_{ik}$'s
give the smallest spacing between their subplanes. $\overline{d}_{ik}^b$
describes the percent change in $\overline{d}_{ik}$ with respect to the bulk
interlayer spacing ($d_b$). 
Note that our calculations are in excellent agreement with 
experiment, except for a slight difference in the lateral shift for the third atom in 
layer two ($p_2^3$), which we find to be negligible and in the opposite 
direction compared to the experimental findings.}

\begin{center}

\begin{tabular}{c|c|c|c}

\hline

Parameters&Our Calculations&Expt. \cite{Schmidt}&Ge et al. \cite{Ge}\\

\hline

$d_b$ (\AA)&1.93&1.92&1.92\\

$d_{12}$ (\AA)&1.96&1.94&1.97\\

$\overline{d}_{12}$ (\AA)&2.24&2.25&-\\

$\overline{d}_{12}^b$ (\%)&16.09&16.67&-\\

$d_{23}$ (\AA)&1.82&1.79&-\\

$\overline{d}_{23}$ (\AA)&1.88&1.88&-\\

$\overline{d}_{23}^b$ (\%)&-2.56&-2.08&-\\

$d_{34}$ (\AA)&1.88&1.83&-\\

$\overline{d}_{34}$ (\AA)&1.94&1.93&-\\

$\overline{d}_{34}^b$ (\%)&0.54&0.52&-\\

$d_{45}$ (\AA)&1.91&1.89&-\\

$\overline{d}_{45}$ (\AA)&1.923&1.91&-\\

$\overline{d}_{45}^b$ (\%)&-0.34&-1.01&-\\

$b_1^{13}$ (\AA)&0.22&0.25&-\\

$b_1^{23}$ (\AA)&0.53&0.55&0.47\\

$b_1^{34}$ (\AA)&0.2&0.20&0.20\\

$p_1^2$ (\AA)&0.03&0.05&0.05\\

$p_1^3$ (\AA)&0.07&0.07&0.02\\

$b_2^{13}$ (\AA)&0.04&0.07&-\\

$b_2^{23}$ (\AA)&0.08&0.10&-\\

$p_2^2$ (\AA)&0.01&0.01&-\\

$p_2^3$ (\AA)&0.0004&-0.02&-\\

$b_3^{13}$ (\AA)&0.06&0.10&-\\

$b_3^{23}$ (\AA)&0.03&0.05&-\\

$b_4^{13}$ (\AA)&0.04&0.06&-\\

$b_4^{23}$ (\AA)&0.01&0.03&-\\

\hline

\end{tabular}

\end{center}

\label{clrecp}

\end{table}

\begin{table}[h]

\caption{Summary of our calculations for the c(2$\times$2) and p(2$\times2$) 
phases and comparison with experiments and previous theoretical calculations (Prev)
\cite{Titmuss}. $E_{ads}$ denotes the adsorption energy of CO 
on Ir\{100\}; $r_{Ir-C}$ and $d_{C-O}$ represent the Ir-C and C-O bond lengths
respectively. $\Delta Z$ gives the vertical distance between two intraplanar 
Ir atoms. $d_{12}$ denotes the distance between the top two surface layers. 
The different parameters are also shown in Fig. \ref{0.5-str}.} 

\begin{center}

\begin{tabular}{|c|c|c|c|c|c|c|c|c|c|c|c|c|c|c|c|}

\hline

\multicolumn{16}{|c|}{c(2$\times$2) phase, 0.5 ML CO coverage}\\

\cline{1-16}

Site & \multicolumn{3}{c|}{$E_{ads}$} &  \multicolumn{3}{c|}{$r_{Ir-C}$} & 

 \multicolumn{3}{c|}{$d_{C-O}$} &  \multicolumn{3}{c|}{$\Delta Z$} &  

\multicolumn{3}{c|}{$d_{12}$}\\

     &  \multicolumn{3}{c|}{eV/ CO molecule} &  \multicolumn{3}{c|}{(\AA)} & 

 \multicolumn{3}{c|}{(\AA)} &  \multicolumn{3}{c|}{(\AA)} & 

 \multicolumn{3}{c|}{(\AA)}\\

\cline{2-16}

   & Ours & Prev & Expt & Ours & Prev & Expt & Ours & Prev & Expt

& Ours & Prev & Expt & Ours & Prev & Expt\\

\hline

Atop & -2.69 & -2.65 & - & 1.86 & 1.86 & 1.81 $\pm$ 0.05 & 1.16 & 1.16 & 1.16

$\pm$ 0.05 & 0.14 & 0.19 & 0.13 $\pm$ 0.05 & 1.82 & 1.82 & 1.82 $\pm$ 0.04\\

Bridge & -2.51 & -2.29 & - & 2.06 & 2.04 & - & 1.17 & 1.18 & - & 0.0 & 0.0 & - &

1.89 & 1.90 &-\\

Hollow& -2.1 & -1.5 & - & 2.29 & 2.28 & - & 1.2 & 1.2 & - & 0.0 & 0.0 & - & 1.90

 & 1.92 &-\\ 

\hline

\multicolumn{16}{|c|}{p(2$\times$2) phase, 0.25 ML CO coverage}\\

\cline{1-16}

Atop & -2.83 & -2.61 & - & 1.86 & 1.86 & - & 1.16 & 1.16 & - & 0.15 & 0.20 & -  

& 1.82 & 1.83 & - \\

Bridge & -2.71 & - & - & 2.05 & - & - & 1.17 & - & - &  0.06 & - & - & 1.84 & -

& -\\

Hollow & -2.34 & - & - & 2.46 & - & - & 1.2 & - & - &  0.0 & - & - &  1.84 & -&

 -\\

\hline

\end{tabular}

\end{center}

\label{compCOse}

\end{table}

\pagebreak

\newpage

\clearpage

\begin{table}[h]

\caption{Adsorption energies ($E_{ads}$) of CO at atop sites on unreconstructed
and reconstructed Ir\{100\} surfaces at different CO coverages. All values of
$E_{ads}$ are expressed in units of eV/CO molecule.}

\begin{center}

\begin{tabular}{c|c|c}

\hline

CO coverage & $E_{ads}$ & $E_{ads}$ \\

(ML)  & unreconstructed surface &reconstructed surface\\

\hline
1.0 & -1.93 & - \\

0.6 & - & -1.90\\

0.5 & -2.55& - \\

0.25 & -2.51 & - \\

0.2 & -2.51 &-2.27\\

0.125 & -2.52& - \\

0.11 & -2.53& - \\

0.10 & - & -2.31 \\

\hline

\end{tabular}

\end{center}

\label{adssur}

\end{table}

\pagebreak

\newpage

\clearpage

\begin{table}[h]

\caption{Adsorption energies ($E_{ads}$) at different sites of the 
reconstructed surface at 0.1 CO coverage. See Figs. \ref{site0.1} and 6 for 
the convention used in labelling sites.}

\begin{center}

\begin{tabular}{cccc}

\hline

coverage & site & $E_{ads}$ \\

(ML) & &(eV/ CO molecule) \\

\hline

    &T1 &  -2.07  \\

    &T2 &  -2.09  \\

    &T3 &  -2.31  \\  

    &T4 &  -1.92  \\

    & H1 &  -1.99   \\ 

    & H2 &  -1.68   \\

    & H3 &  -1.70   \\ 

    & H4 &  -2.01   \\ 

 0.1&H5 &  -2.21   \\ 

    & H6 &  -1.89   \\ 

    & B1&  -2.00 \\

    & B2&  -1.62 \\

    & B3&  -2.19 \\

    & B4&  -1.63 \\

    & D1 &  -1.76  \\

    & D2 &  -1.73  \\

    & D3 &  -2.00  \\

\hline

    & A and B &-2.27\\

0.2 & A and C &-2.27\\

    & A and D &-2.15\\

\hline

0.6& A, B, E, F, G, H &-1.90\\

\hline

\end{tabular}

\end{center}

\label{CO_rec_ads}

\end{table}

\pagebreak

\newpage

\clearpage

\pagebreak

\newpage

\clearpage

\newpage

\clearpage

\newpage

\clearpage

\begin{figure}[p]

\centering

\includegraphics[scale=0.60]{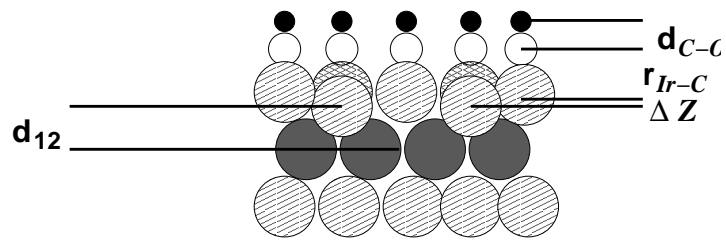}

\caption{Schematic side view of the structure of CO adsorbed on 
unreconstructed Ir\{100\}-(2$\times$2) at 0.5 ML coverage\cite{Titmuss}. The
big, medium and small circles represent Ir, O and C atoms.}

\label{0.5-str}

\end{figure}

\newpage

\clearpage

\newpage

\clearpage

\newpage

\clearpage

\newpage

\clearpage

\newpage

\clearpage

\end{document}